\documentclass[prd,showpacs]{revtex4}
\usepackage{epsfig}
\usepackage{amsmath}
\usepackage{amssymb}

\begin{document}

\title{ Nucleon electromagnetic form factors in a quark-gluon core model}

\author{Xian-Qiao Yu\footnote{yuxq@swu.edu.cn}}

\affiliation
 {\it \small School of Physical Science and Technology, Southwest University, Chongqing 400715, China}

\begin{abstract}

We study the nucleon electromagnetic form factors in a quark-gluon
core model framework, which can be viewed as an extension of the
Isgur-Karl model of baryons. Using this picture we derive nucleon
electromagnetic dipole form factors at low $Q^2$ and the deviation
from the dipole form at high $Q^2$, that are consistent with the
existing experimental data.

\end{abstract}

 \maketitle

 The nucleon elastic electromagnetic form factors(FFs) are very
important for understanding the dynamics of the nucleons'
constituents. There has been much activity in the measurement of
proton and neutron elastic electromagnetic FFs in the last decade.
 High accuracy experimental data on nucleon
electromagnetic FFs
 obtained in recent years\cite{Arrington,Qattan,Kubon,Brooks} indicate that nucleon electromagnetic
 FFs can be well fitted by a simple dipole formula at low $Q^2$:

\begin{equation}
 G_{E}^{p}=G_{M}^{p}/\mu_{p}=G_{M}^{n}/\mu_{n}=1/(1+Q^2/0.71GeV^2)^2,\label{dipoleex}
\end{equation}
here $Q^2=-q^2$, $q$ is momentum transfer in elastic
electron-nucleon scattering and $\mu_{p}$ and $\mu_{n}$ are
magnetic moments of proton and neutron respectively. This has
spurred a significant reevaluation of the nucleon and pictures of
its structure\cite{PPV}.

Starting from the relation between nucleon electromagnetic FFs and
nucleon intrinsic charge(magnetic moment) density distributions,
this note will give a possible origin of the dipole FF. Strict
analysis of the dynamics of nucleon constituents should start from
QCD. Because the complication of QCD non-perturbation, various
phenomenological models are developed. Quark potential
model\cite{LS,HK,RCI} treats the nucleon as three constituent
quarks bound state, the effect of gluons is buried within
constituent quarks, which are considered as quasi-particles.
Hadrons bag model\cite{Bog,Chodos,Degrand,CT} assumes the
nucleons' three non-interacting quarks are confined in a bag of
finite dimension. These models have absorbed or ignored other
degrees of freedom beyond the three quarks. If we take into
account the gluon degree of freedom in nucleon and assume gluons
in nucleon contract under their own strong self-interactions, we
find a different method of describing nucleon from which the
dipole FF can be derived easily.

\begin{figure}[htb]
  \begin{center}
  \epsfig{file=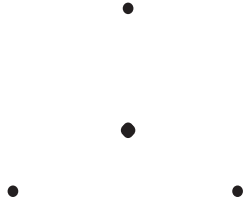,width=100pt,height=80pt}
   \end{center}
   \caption{The configuration for quark-gluon core structure model of nucleon}
   \label{figure:Fig1}
  \end{figure}

The idea of this picture about nucleon structure comes from a
comparison between nucleon and triatomic molecule. We suppose a
quark at a place in space, unlike its electric charge located at
the definite place where it is, its color charge will diffusely
spread out around it due to gluon emission and absorption, this
leads to that most of the quark's color charge is carried by the
gluon cloud around it. Comparing nucleon with triatomic molecule,
three valence quarks corresponding approximately to three atomic
nuclei, the gluon cloud around valence quark is just like electron
cloud around atomic nucleus. In the central core of three valence
quarks, gluon cloud will over lap and become dense.  We may assume
that the dense gluon cloud will contract under its own strong
interaction to a compact gluon cluster( Boros $et$ $al$. have
suggested virtual gluon clusters exist in nucleon in
refrence\cite{Boros}, where gluon clusters mean a group of
gluons), three light valence quarks, part of their color charge
have been transferred to the compact gluon cluster, moving around
the nucleus composed of gluons. The spin-independent interaction
between one valence quark and the gluon nucleus take the following
form:
\begin{equation}
  V=-\frac{\alpha}{r}+V_{conf},\label{potential}
\end{equation}
where $\alpha$ is a positive constant and $V_{conf}$ is the
confining potential. We call this picture quark-gluon core
structure model of nucleon, as shown in Fig.~\ref{figure:Fig1}.

 We find that the above hypothesis combined with quantum mechanics
appears sufficient for the derivation of nucleon electromagnetic
dipole FFs at low $Q^2$. In the following I shall sketch the
derivation briefly.

\begin{figure}[htb]
  \begin{center}
  \epsfig{file=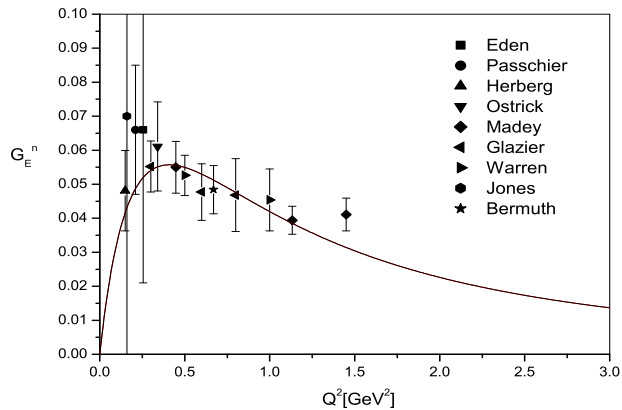,width=260pt,height=180pt}
   \end{center}
   \caption{The neutron electric form factor $G^{n}_{E}$ as a function of $Q^{2}$. Data are
    from references\cite{Eden,Passchier,Herberg,Ostrick,Madey,Glazier,Warren,Jones,Bermuth}.}
   \label{figure:Fig2}
  \end{figure}

We assume that the mass of the gluon cluster is much larger than that of the quark, in this case,
setting the gluon nucleus at the origin of coordinate, we
write the non-relativistic Hamiltonian for the system as

\begin{equation}
 H=H_{0}+H_{pert},\label{hami}
\end{equation}
where
\begin{equation}
 H_{0}=-\frac{\hbar^{2}}{2m}\nabla_{1}^{2}-\frac{\hbar^{2}}{2m}\nabla_{2}^{2}-\frac{\hbar^{2}}{2m}\nabla_{3}^{2}
 -\frac{\alpha}{r_{1}} -\frac{\alpha}{r_{2}}
 -\frac{\alpha}{r_{3}},\label{hami0}
\end{equation}
and
\begin{equation}
 H_{pert}=-\frac{\beta}{r_{12}}-\frac{\beta}{r_{13}}-\frac{\beta}{r_{23}}+\sum_{i<j}(V_{conf}^{ij}+V_{hyp}^{ij}),\label{hamip}
\end{equation}

where $V_{hyp}$ is the hyperfine interaction which is
spin-dependent. The last term in Eq.(\ref{hamip}) includes all the
confining and hyperfine interactions among the quarks and gluon
core. The hyperfine interaction between each pair of quarks (i, j)
given by\cite{Rujula} is
\begin{equation}
 V_{hyp}^{ij}=\frac{2\alpha_{s}}{3m_{i}m_{j}}\left[\frac{8\pi}{3}\vec{S_i} \cdot\vec{S_j}\delta^{3}(\vec{r}_{ij})
 +\frac{1}{r_{ij}^{3}}\left(\frac{3\vec{S_i}\cdot \vec{r}_{ij}\vec{S_j}\cdot\vec{r}_{ij}}{r_{ij}^{2}}-
 \vec{S_i}\cdot\vec{S_j} \right) \right],\label{vhyp}
\end{equation}
but its form between the valence quark and the gluon core is
unknown. If we take a harmonic oscillator type confining
potential, the picture described here can be viewed as an
extension of the Isgur-Karl model\cite{Isgur}. We can use it to
study the masses of baryons, which I hope to discuss in a separate
paper. Here we are interested in nucleon electromagnetic form
factors. The eigenstates of the Hamiltonian(\ref{hami0}) are well
known. For nucleon, all the three valence quarks in the $1s$
state, the ground state wave function is

\begin{equation}
  \psi_{0}(r_{1},r_{2},r_{3})=\psi_{100}(r_1)\psi_{100}(r_2)\psi_{100}(r_3)=\frac{b_{0}^{9/2}}{\pi\sqrt{\pi}}\exp\left[-b_{0}(r_1+r_2+r_3)\right],
\label{wave}
\end{equation}
where $b_{0}=\alpha m/\hbar^{2}$. It is impossible to solve
accurately the eigen-wave functions of Hamiltonian(\ref{hami});
 we assume that
the approximate ground state wave function of
Hamiltonian(\ref{hami}) has the same form as Eq.(\ref{wave}), that
is
\begin{equation}
  \psi(r_{1},r_{2},r_{3})=\frac{b^{9/2}}{\pi\sqrt{\pi}}\exp\left[-b(r_1+r_2+r_3)\right],
\label{wavem}
\end{equation}
where $b=\alpha' m/\hbar^{2}$ and $\alpha'$ is a effective coupling constant. A direct result of Eq.(\ref{wavem})
is that the electric charge and magnetic moment density
distributions in nucleon are a function of exponential type

\begin{equation}
  \rho(r)=\frac{b^3}{\pi}\exp[-2br].
\label{dist}
\end{equation}
After carrying through Fourier transformation

\begin{equation}
  F(q^2)=\frac{4\pi}{q}\int  \rho(r)\sin(qr)rdr,
  \label{Ftrans}
\end{equation}
we get nucleon electromagnetic form factors

\begin{gather}
 G^{p}_{E}(Q^2) = \frac{ G^{p}_{M}(Q^2)}{\mu_{p}}=\frac{
 G^{n}_{M}(Q^2)}{\mu_{n}}\equiv G_{D}(Q^2),\\
 G^{n}_{E}(Q^2) =0,\label{formf}
\end{gather}
where
\begin{equation}
G_{D}(Q^2)= \frac{1}{\left(1+Q^2/4b^2\right)^{2}},\label{dipole}
\end{equation}
which is dipole formula supported by many
experiments\cite{Arrington,Qattan,Kubon,Brooks}. We notice that a
small but definite deviation from zero is observed for the
        neutron electric form factor $G^{n}_{E}$ at low $Q^2$ by recent high-precision
        data from double-polarization
        measurements\cite{Hyde-Wright}, which can be explained by
        the small mass and interaction differences between
        up quark and down quark in our model. Look back to
        Eq.(\ref{hami0}), where we have assumed up quark and
down quark to have the same mass and coupling constant. Denoting
$m_{u}$ and $m_{d}$ as the mass of up quark and down quark,
respectively, $\alpha'_{u}$ and $\alpha'_{d}$ represent
respectively the effective coupling constant involving of $u$ and
$d$ quark, and we have the following expression for the
        neutron electric form factor

\begin{equation}
G^{n}_{E}(Q^2)
=\frac{2}{3}\left[\frac{1}{\left(1+Q^2/4b_{u}^2\right)^{2}}-\frac{1}{\left(1+Q^2/4b_{d}^2\right)^{2}}\right],\label{gen}
\end{equation}
where  $b_{u}=\alpha'_{u} m_{u}/\hbar^{2}$ and $b_{d}=\alpha'_{d}
m_{d}/\hbar^{2}$. If we assume that $m_{u}=m_{d}=m$ and
$\alpha'_{u}=\alpha'_{d}=\alpha'$, we find that constraints from
dipole formula and energy spectra of nucleons suggest
$\alpha'=3.17$ and $m =133MeV$. Supposing that the mass difference
between
        up quark and down quark is $7MeV$(we set $m_{d} =133MeV$ and $m_{u} =140MeV$) and the effective coupling
        constant symmetry breaking is about $9\%$(we set $\alpha'_{d}=3.17$ and $\alpha'_{u}=3.47$), we get
        the neutron electric form factor $G^{n}_{E}$ as shown in
Fig.~\ref{figure:Fig2}, which is consistent with the existing experimental data\cite{Hyde-Wright}.

In the above discussions, we interpret the Fourier transforms of nucleon charge(magnetization)
densities as the electromagnetic FFs. This identification is only appropriate for a non-relativistic(static) system.
 However, if the wavelength of the probe is much
 larger than the Compton wavelength of the nucleon with mass $M_{N}$, i.e. if $|Q^{2}|\gg M_{N}^{2}$, one needs to take the
 effect of relativity into account and consequently the physical interpretation of the FFs becomes complicated. Recently, Kelly\cite{Kelly} has
 used a relativistic prescription to relate the electromagnetic FFs to the nucleon charge and magnetization densities, accounting for
 the Lorentz contraction of the densities in the Breit frame relative to the rest frame. We follow this treatment in reference\cite{Kelly}
 to give the nucleon electromagnetic FFs at high $Q^{2}$.

 Let $\rho_{ch}(r)$ and $\rho_{m}(r)$ represent the spherical charge and magnetization densities in the nucleon rest frame, the related
 intrinsic FFs can be obtained through a Fourier-Bessel transform as\cite{Kelly}
\begin{equation}
\widetilde{\rho}(k)=\int_{0}^{\infty}dr r^{2}j_{0}(kr)\rho(r),\label{fbt}
\end{equation}
with $k\equiv |q|$ being the wave vector in the nucleon rest frame. At low $Q^{2}$, the nucleon is a non-relativistic system and the intrinsic FFs
are just the the electromagnetic FFs $G_{E}(Q^{2})$ and $G_{M}(Q^2)$(called Sachs FFs in the literature)
 that have been discussed above. However, at high $Q^2$ where the nucleon moves with velocity $v=\sqrt{\tau/(1+\tau)}$ relative to
 the rest frame, here $\tau=Q^{2}/4M_{N}$, a Lorentz boost with $\gamma^{2}=(1-v^{2})^{-1}=1+\tau$ is involved\cite{PPV}.
  This Lorentz boost leads to a contraction of the nucleon
 densities as seen in the Breit frame and hence the intrinsic FFs defined by Eq.(\ref{fbt}) needs to replace $k^2$ with $Q^2/(1+\tau)$.

\begin{figure}[htb]
  \begin{center}
  \epsfig{file=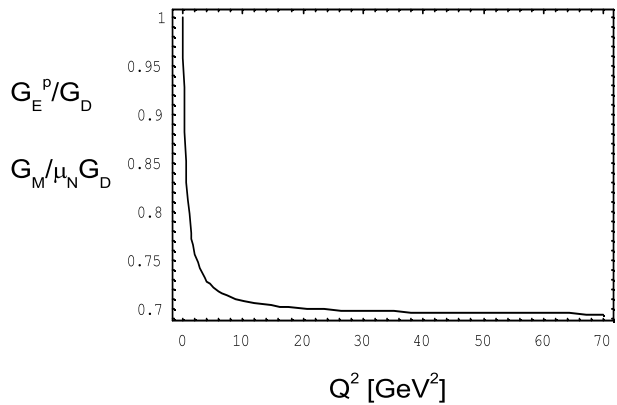,width=260pt,height=180pt}
   \end{center}
   \caption{The proton electric form factor $G_{E}^{p}$ in unit of $G_{D}$ and the nucleon
   magnetic form factor $G_{M}$ in units of $\mu_{N}G_{D}$ as a function of $Q^2$.}
   \label{figure:Fig3}
  \end{figure}

The relativistic relationships between the intrinsic FFs $\widetilde{\rho}(k)$ and the Sachs FFs $G(Q^2)$ measured
by electron scattering at finite $Q^2$ are not unambiguous. There exist different prescriptions in the literature which
can be written in the form

\begin{gather}
 \widetilde{\rho}_{ch}(k)=G_{E}(Q^2)(1+\tau)^{\lambda_{E}},\label{rel1} \\
 \mu_{N}\widetilde{\rho}_{m}(k)=G_{M}(Q^2)(1+\tau)^{\lambda_{M}},\label{rel2}
\end{gather}
where $\lambda$ is a model-dependent constant. To account for the
asymptotic $1/Q^4$ FFs obtained by the perturbative QCD at very
large $Q^2$, Mitra and Kumari\cite{Mitra} proposed the choice
$\lambda_{E}=\lambda_{M}=2$. Following this choice we calculate
the nucleon electromagnetic FFs at very high $Q^2$ as shown in
Fig.~\ref{figure:Fig3}.

  From Fig.~\ref{figure:Fig3} we can see that $G_{E}^{p}/G_{D}=G_{M}/\mu_{N}G_{D}\approx0.7$ at high $Q^2$, which
  is consistent with the existing experimental data for $G_{M}^{p}/\mu_{p}G_{D}$
   in the range of momentum transfer from $Q^2=19.5$ to 31.3 $(GeV/c)^2$\cite{Sill}. The experimental data at
   higher $Q^2$ values is not available nowadays and will become available in the near future that will provide
    a critical test of our calculations.

    If we choose $\lambda_{M}^{p}=1.9$ and $\lambda_{M}^{n}=2$, we
    obtain $\frac{1}{\mu_{p}}Q^{4}G_{M}^{p}(Q^2)\approx 0.4 GeV^4$
    and $\frac{1}{\mu_{n}}Q^{4}G_{M}^{n}(Q^2)\approx 0.3 GeV^4$ at
    $Q^{2}\simeq10-30 GeV^{2}$, that are consistent with the
    perturbative QCD results calculated by Chernyak and Zhitnitsky \cite{CZ}.
    They also predicted that
    $\frac{1}{\mu_{p,n}}Q^{4}G_{M}^{p,n}(Q^{2})\rightarrow 0$ at
    $Q^{2}\rightarrow \infty$\cite{CZ}, which disagrees with the
    predictions of Eq.(\ref{rel2}) for the
    choice $\lambda_{M}^{p}=1.9$ and $\lambda_{M}^{n}=2$.
    Considering that the unique relativistic relationships between the intrinsic FFs $\widetilde{\rho}(k)$ and the Sachs FFs $G(Q^2)$ do not exist,
   there might be large uncertainties in Eq.(\ref{rel2}) at
   $Q^{2}\rightarrow\infty$. Eventually the uncertainties can be extracted
   in the future form factor measurements at higher $Q^2$ values.

We consider the gluon core inside nucleon as a quasi-particle.
Further, studies along this line will show the properties of such
quasi-particle, for example its spin. These properties are helpful
information for our quantitative understanding of nucleon.

The author thanks M. Z. Zhou for useful discussions. This work is
partly supported by the National Natural Science Foundation of
 China under Grant No. 10847157.


\begin{thebibliography}{99}
\bibitem{Arrington}
J. Arrington, Phys. Rev. C{\bf69}, 022201(2004).
\bibitem{Qattan}
I. A. Qattan, $et$ $al$, Phys. Rev. Lett{\bf94}, 142301(2005).
\bibitem{Kubon}
G. Kubon, $et$ $al$, Phys. Lett. B{\bf524}, 26(2002).
\bibitem{Brooks}
W. K. Brooks and J. D. Lachniet for the CLAS Collaboration, Nucl.
Phys. A{\bf755}, 261(2005).
\bibitem{PPV}
See two review articles: C. F. Perdrisat, V. Punjabi, M.
Vanderhaeghen, Prog. Part. Nucl. Phys.{\bf59}, 694(2007) and
references therein; J. Arrington, C. D. Roberts, J. M. Zanotti, J. Phys. G.{\bf34}, S23(2007) and references therein.
\bibitem{LS}
 W. Lucha, F. F. Sch\"oberl, D. Gromes, Phys. Rep., {\bf 200},
127(1991).
\bibitem{HK}
 A. J. G. Hey, R. L. Kelly, Phys. Rep., {\bf96},
71(1983).
\bibitem{RCI}
 J. M. Richard, Phys. Rep., {\bf212}, 1(1992); S. Capstick, N. Isgur. Phys. Rev. D{\bf34}, 2809(1986).
 \bibitem{Bog}
 P. N. Bogolioubov, Ann. Inst. Henri Poincare, {\bf8}, 163(1968).
\bibitem{Chodos}
  A. Chodos, $et$ $al$, Phys. Rev. D{\bf9}, 3471(1974); D{\bf10}, 2599(1974).
 \bibitem{Degrand}
  T. DeGrand, $et$ $al$, Phys. Rev. D{\bf12}, 2060(1975).
 \bibitem{CT}
 A. Chodos, C. B. Thorn, Nucl. Phys. B{\bf104}, 21(1976).
\bibitem{Boros}
C. Boros, Liang Zuo-tang and Meng Ta-chung, Phys. Rev. D{\bf54},
6658(1996).
\bibitem{Rujula}
A. De. Rujula, H. Georgi and S. L. Glashow,  Phys. Rev. D{\bf12},
147(1975).
\bibitem{Isgur}
N. Isgur and G. Karl,  Phys. Rev. D{\bf18}, 4187(1978); {\bf19}, 2653(1979); {\bf20}, 1191(1979);
{\bf23}, 817(1981)(Errata).
\bibitem{Hyde-Wright}
for a review, see, C. E. Hyde-Wright and K. de Jager, Annu. Rev.
Nucl. Part. Sci. {\bf54}, 217(2004).
\bibitem{Eden}
T. Eden, $et$ $al$, Phys. Rev. C{\bf50}, R1749(1994).
\bibitem{Passchier}
I. Passchier, $et$ $al$,  Phys. Rev. Lett{\bf82}, 4988(1999).
\bibitem{Herberg}
C. Herberg, $et$ $al$, Eur. Phys. Jour. A{\bf5}, 131(1999).
\bibitem{Ostrick}
M. Ostrick, $et$ $al$, Phys. Rev. Lett{\bf83}, 276(1999).
\bibitem{Madey}
R. Madey, $et$ $al$, Phys. Rev. Lett{\bf91}, 122002(2003).
\bibitem{Glazier}
D. I. Glazier, $et$ $al$, Eur. Phys. Jour. A{\bf24}, 101(2005).
\bibitem{Warren}
G. Warren, $et$ $al$, Phys. Rev. Lett{\bf92}, 042301(2004); H. Zhu,
 $et$ $al$, Phys. Rev. Lett{\bf87}, 081801(2001).
\bibitem{Jones}
C. E. Jones-Woodward, $et$ $al$, Phys. Rev. C{\bf44}, R571(1991).
\bibitem{Bermuth}
J. Bermuth, $et$ $al$, Phys. Lett. B{\bf564}, 199(2003); D. Rohe,
$et$ $al$, Phys. Rev. Lett{\bf83}, 4257(1999).
\bibitem{Kelly}
J. J. Kelly, Phys. Rev. C{\bf66}, 065203(2002).
\bibitem{Mitra}
A. N. Mitra and I. Kumari, Phys. Rev. D{\bf15}, 261(1977).
\bibitem{Sill}
A. F. Sill,  $et$ $al$, Phys. Rev. D{\bf48}, 29(1993).
\bibitem{CZ}
V. L. Chernyak and I. R. Zhitnitsky, Nucl. Phys. B{\bf 246},
52(1984); Phys. Rep. {\bf 112}, 173(1984).

\end{thebibliography}
\end{document}